\documentclass[superscriptaddress,twocolumn,prl]{revtex4-1}
\usepackage{amsmath}
\usepackage{amssymb}
\usepackage{graphicx}
\usepackage{dcolumn}
\usepackage{bm}

\usepackage{bm}
\usepackage{graphicx}
\usepackage{ulem}

\def\be{\begin{equation}}
\def\ee{\end{equation}}
\def\ber{\begin{eqnarray}}
\def\eer{\end{eqnarray}}
\def\bwt{\begin{widetext}}
\def\ewt{\end{widetext}}

\def\nn{\nonumber}

\def\e{{\varepsilon}}

\def\e{\varepsilon}
\def\o{\omega}

\begin{document}

\newcommand{\sgn}{\mathop{\mathrm{sgn}}}
\newcommand{\ef}{\mathop{\varepsilon_{F}}}

\def\o{\omega}
\def\e{\varepsilon}
\def\nn{\nonumber}

\title{Piezoelectric surface acoustical phonon limited mobility of electrons\\ in graphene on a GaAs substrate}

\author{S. H. Zhang}
\affiliation{Department of Physics, University of Antwerp, Groenenborgerlaan 171, B-2020 Antwerpen, Belgium}
\affiliation{Key Laboratory of Materials Physics, Institute of Solid State Physics, Chinese Academy of Sciences, Hefei 230031, China}
\author{W. Xu}
\affiliation{Key Laboratory of Materials Physics, Institute of Solid State Physics, Chinese Academy of Sciences, Hefei 230031, China}
\author{S. M. Badalyan}
\email{Samvel.Badalyan@ua.ac.be}
\affiliation{Department of Physics, University of Antwerp, Groenenborgerlaan 171, B-2020 Antwerpen, Belgium}
\author{F. M. Peeters}
\affiliation{Department of Physics, University of Antwerp, Groenenborgerlaan 171, B-2020 Antwerpen, Belgium}

\begin{abstract}
We study the mobility of Dirac fermions in monolayer graphene on a GaAs substrate, restricted by the combined action of the extrinsic potential of piezoelectric surface acoustical phonons of GaAs (PA) and of the intrinsic deformation potential of acoustical eigen-phonons in graphene (DA). In the high temperature ($T$) regime the momentum relaxation rate exhibits the same linear dependence on $T$ but different dependences on the carrier density $n$, corresponding to the mobility $\mu\propto 1/\sqrt{n}$ and $1/n$, respectively for the PA and DA scattering mechanisms. In the low $T$ Bloch-Gr\"uneisen regime, the mobility shows the same square-root density dependence, $\mu\propto \sqrt{n}$, but different temperature dependences, $\mu\propto T^{-3}$ and $ T^{-4}$, respectively for PA and DA phonon scattering.
\end{abstract}

\maketitle


Graphene \cite{NovGeim2004} due to its unique linear chiral electronic dispersion \cite{Castro2009} exhibits novel  transport properties \cite{NovGeim2005,Tan2007} and has great potential as a desirable material for future electronic and optical technologies \cite{GeimAMD2007,Castro2009}. 
Momentum relaxation is a key phenomenon that governs transport of Dirac fermions in graphene \cite{DaSarma2011}. It is of practical interest for developing high-speed electronics and in recent years has been extensively studied both theoretically \cite{Katsnelson2008,Hwang2011,Kaasbjerg2012} and experimentally \cite{Bolotin2008,Efetov2010,Yan2011}. The scattering by defects \cite{Chen2009,Katsnelson2008,Morozov2008,Ishigami2007,Stauber2007}, impurities \cite{Nomura20067,Ando2007,Hwang2007L,Hwang2011,Yan2011,ChenNP2008,Hwang2009}, and phonons 
\cite{Stauber2007,Hwang2007,Chen2008,Zou2010,Efetov2010,Min2011,Zhang2011,Kaasbjerg2012} have been investigated to determine and control the dominant mechanism that limits the carrier mobility in graphene.

In device structures used so far, graphene is often deposited on an oxidized silicon wafer (SiO$_{2}$/Si), which due to various scattering mechanisms imposes constraints on its excellent transport properties observed in suspended graphene devices \cite{Bolotin2008,Du2008}. Recently, structures on other promising substrate materials such as h-BN \cite{Decker2011,Dean2011} and GaAs \cite{Ding2010,Woszczyna2012} have been fabricated and studied with the intention for high-quality graphene electronics. Along with its superior surface quality and strong hydrophilicity preventing folding of large-scale graphene flakes, GaAs has a substantially larger dielectric constant in comparison with SiO$_{2}$ and h-BN and hence improved electrical screening. In such high purity GaAs structures, electron-phonon scattering can be a decisive factor in limiting the mobility of Dirac fermions and the piezoelectric GaAs substrate can serve as a powerful tool for studying electronic properties of graphene by means of remote piezoelectric surface acoustical phonons.

In the present work we study the temperature and density dependence of the carrier mobility in monolayer graphene at finite doping on a GaAs substrate. We calculate the mobility limited by scattering from the piezoelectric potential of remote surface acoustical phonons of the substrate (PA phonons) versus the deformation potential of acoustical eigen-phonons of the graphene lattice (DA phonons). In experiment the typical wavelength of phonons taking part in scattering events is much larger than the distance, $d$, of  several angstr\"oms between the graphene sheet and the GaAs substrate so that extrinsic interaction of remote PA phonons with Dirac fermions is quite strong and, as we shall see, can dominate the intrinsic interaction with the DA phonons of graphene. 

Crystal surfaces modify substantially bulk phonon modes and can change qualitatively the carrier relaxation characteristics \cite{SMB1988}. In crystals with lack of a center of symmetry such as in GaAs, the displacement field of Rayleigh waves (a combination of longitudinal and transversal oscillations) propagating on a crystal free surface \cite{Landau} induces a piezoelectric polarization of the lattice. It leads to an electric potential both inside and outside of the GaAs substrate that couples to the electrons in graphene.
%
%
The Hamiltonian of such piezoelectric interaction between surface acoustical phonons of the GaAs substrate and the massless Dirac fermions in graphene can be written as
\begin{eqnarray}\label{PA}
H^{PA}_{e-ph}=e \varphi({\bf R},t)={1\over\sqrt{\cal{A}}}\sum_{{\bf q}} \gamma^{PA}_{q} e^{i {\bf q R } }b_{\bf q}+\text{c.c.}
\end{eqnarray}
where the electric potential $\varphi({\bf R},t)$ in the plane of the graphene sheet is given by the solution of the Poisson equation as $\varphi({\bf R},t) \propto \hat{q}_x\hat{q}_y e^{i \left( {\bf q R }-i \o_{q} t \right)}e^{-q d}$  \cite{Levinson1996}. Here 
$q$ and $\o_{q}=v_{PA}q$ are the phonon momentum and energy with the velocity of surface Rayleigh waves $v_{PA}\approx 0.9 v_{b}\approx 2.7 \times10^3$ m/s, $v_{b}$ is the bulk sound velocity in GaAs and $\hat{q}_{x,y}=q_{x,y}/q$. 
In Eq.~(\ref{PA}) $b_{\bf q}$ denotes the amplitude of the phonon field and the piezoelectric electron-phonon interaction vertex is defined as $\left|\gamma^{PA}_{q}\right|^{2}=c_{PA}^{2} \left(\hat{q}_{x}\hat{q}_{y}\right)^{2} e^{-2 q d} \hbar^{2}v_{PA}/\left(p_{0}{\bar\tau}_{PA}\right)$
where we use the nominal time ${\bar \tau}_{PA} \approx 8$ ps, introduced in Ref.~\onlinecite{GL} for carrier scattering from bulk piezoelectric acoustical phonons. Here the characteristic wave vector $p_0=2.5\times10^6$ cm$^{-1}$ is related to the optical phonon energy in GaAs and the numerical factor $c_{PA}\approx 4.9$ is determined by the elastic properties of GaAs \cite{Levinson1996}.  The strongest electron-PA phonon interaction takes place for surface phonons propagating along the diagonal direction with $q_{x}\approx q_{y}$ so that in (\ref{PA}) we can approximate $\left(\hat{q}_{x} \hat{q}_{y}\right)^{2}\approx 1/4$. Taking also $e^{-2 q d}\approx 1$ for typical values of $d\sim 5$ \AA~ \cite{Ding2010,Woszczyna2012}, one can see that after these well justified simplifications, the PA phonon vertex $\gamma_{q}^{PA}$ becomes independent of the phonon momentum $q$. This differs from the linear wave vector dependence of the DA interaction vertex and results in a new, qualitatively different, contribution to the mobility.

The Hamiltonian of electron-acoustical phonon interaction due to the deformation potential in graphene can be written as 
\begin{eqnarray}
H_{DA}={1\over\sqrt{\cal{A}}}\sum_{{\bf q}} \gamma^{DA}_{q} e^{i {\bf q R } }b_{\bf q}+\text{c.c.}
\end{eqnarray}
with the DA interaction vertex defined as $\left|\gamma_q^{DA}\right |^{2}=\hbar^2 q v_{DA} / \left(p_0^2 \bar{\tau}_{DA}\right)$.
Here we introduce the nominal scattering time $1/\bar{\tau}_{DA}= \Xi^{2} p_0^2 / \left(2 \hbar \rho v^{2}_{DA}\right)$ for DA phonon interaction and find $\bar{\tau}_{DA}\approx 0.8$ ps using the following values for the graphene parameters \cite{Kaasbjerg2012}: $\Xi=6.8$ eV the coupling constant of the deformation potential, $\rho=7.6 \times 10^{-7}$ kg m$^{-2}$ the surface mass density of graphene, and $v_{DA}=2.0\times10^4$ m/s the sound velocity in graphene, which is essentially larger from the sound velocity in GaAs. The direct  comparison $\left|\gamma_q^{PA}\right|^2 / \left|\gamma^{DA}_{q}\right|^{2}\approx 2.0\times10^{5}~\text{cm}^{-1}/q$ shows that both the PA and DA electron-phonon scattering mechanisms can be important in typical experimental situations and their actual contributions to the graphene mobility are determined by the carrier density, $n$, and temperature, $T$. 

%
The momentum relaxation rate of a test electron due to scattering from the potential of the $s=$ PA, DA phonon field is
\begin{eqnarray}\label{MRR}
{1\over\tau^{s}_{1}(\varepsilon_{\bf \lambda k})}=\sum^{\pm}_{\lambda' {\bf k}'; s {\bf q}}(1-\cos\theta_{{\bf k}{\bf k}'})W^{\pm s {\bf q}}_{\lambda {\bf k}\rightarrow \lambda' {\bf k}'}{1-f(\varepsilon_{\lambda' \bf k'})\over 1- f(\varepsilon_{\lambda \bf k})}
\end{eqnarray}
where $\e_{\lambda \bf k}=\lambda v_{F} k$ is the electron energy in monolayer graphene with the chirality $\lambda$ and the Fermi velocity $v_{F}$. The Fermi functions $f(\e_{\lambda \bf k})$ are determined by the electron Fermi energy $\e_{F}$ and the lattice temperature $T$. The electron transition probability due to the emission ($+$) and absorption ($-$) of phonons is 
\begin{eqnarray}\label{TP}
W^{\pm s {\bf q}}_{\lambda {\bf k}\rightarrow \lambda' {\bf k}'}={2\pi\over\hbar} \left| M^{\pm s {\bf q}}_{\lambda {\bf
k}\rightarrow \lambda' {\bf k}'}\right|^{2} \left( N\left(\omega_{s {\bf q}}\right) + \frac{1}{2} \pm \frac{1}{2} \right) \\
\times
\delta\left(\varepsilon_{\lambda \bf k}-\varepsilon_{\lambda' \bf k'} \mp \hbar \omega_{s {\bf q}}\right) \nn
\end{eqnarray}
with the Bose factors $N\left(\omega_{s {\bf q}}\right)$ representing the number of phonons of the $s, {\bf q}$ mode. 
%
%
Making use of the electron wave functions in graphene $\psi^{T}({\bf R})=\left|e^{-i\theta_{\bf k}},\lambda \right| e^{-i {\bf k \cdot R}}/\sqrt{2\cal{A}}$ where $\theta_{\bf k}$ is the polar angle of the electron wave vector ${\bf k}$, we obtain for the square modulus of the PA and DA matrix elements
\begin{eqnarray}
\left|M^{\pm s {\bf q}}_{\lambda {\bf k}\rightarrow \lambda' {\bf k}'}\right|^{2}=\frac{\delta_{\bf k', k \mp q}}{\cal{A}} \left|\gamma^{s}_{q}\right|^{2} {\cal F}_{\lambda'\lambda}(\theta_{{\bf k}{\bf k}'})~.
\end{eqnarray}
Here $\theta_{\bf k k'}=\theta_{\bf k'}- \theta_{\bf k}$ and the form factor ${\cal F}_{\lambda' \lambda}(\theta_{\bf k k'})=\left(1+\lambda\lambda'\cos \theta_{\bf k k'} \right)/2$ represents an overlap of the electron spinor wave functions. 
We consider electron-phonon scattering in doped graphene with $\e_{F}$ much larger than the typical acoustical phonon energy. In such samples, only intra-chirality subband transitions are effective so we take $\lambda=\lambda'=1$ and further omit this index. Then, we can represent the relaxation rate due to extrinsic PA and intrinsic DA phonon scattering  as
\begin{eqnarray}\label{MRREX}
{1\over\tau^{PA,DA}_{1}(\e_{\bf k})}=\frac{a c^{2}}{\pi}\frac{1}{{\bar\tau}_{PA,DA}} \left(\frac{k}{p_{0}}\right)^{1+m} {\cal G}_{2+m}(x)
\end{eqnarray}
where we introduce the functions
\begin{eqnarray}\label{MRRF}
{\cal G}_{k}(x)= \sum_{\pm}\int^{z_{\pm}}_{0} dz z^{k} \eta(a,z) \Psi^{\pm}(x,y)
\end{eqnarray}
with  $z=q/2k$ and $z_{\pm}= 1/(1\pm a)$. In Eq.~(\ref{MRREX}) for PA phonons we take $m=0$ and $a=a_{PA}=v_{PA}/v_{F}$, $c=c_{PA}$; for DA phonons $m=1$, $a=a_{DA}=v_{DA}/v_{F}$, $c=c_{DA}=2\sqrt{2}$.
For brevity, we define also the functions $\Psi^{\pm}(x,y)=\left( N(y) +1/2 \pm 1/2 \right) \left(1-f(x \mp y)\right)/ \left(1-f(x)\right)$
where $x= (\e_{ \bf k}-\e_{F})/T$ and $y = \omega_{s {\bf q}}/T=z(k/k_F) (T^{PA,DA}_{BG}/T)$. The characteristic  Bloch-Gr\"uneisen (BG) temperatures $T^{PA, DA}_{BG}=2\hbar v_{PA, DA}k_{F}$ are defined, respectively, for PA and DA phonon scattering.
In Eq.~(\ref{MRRF}) the function $\eta(a,z)=\sqrt{1-\left(a \pm \left(1-a^2\right) z\right)^2}/ (1 \mp 2 a z )$ appears due to the chiral nature of Dirac fermions and restricts backscattering processes in graphene. As far as the Fermi velocity in graphene $v_{F}\approx 1.15\times 10^{6}$ m/s is much larger than the sound velocity in GaAs and in graphene, in our analytical calculations we take $\eta(a,z)\approx\eta(0,z)=\sqrt{1-z^{2}}$ and $z_{\pm}\approx 1$.

First we analyze analytically the momentum relaxation of a test electron in the regimes where PA and DA phonon scattering are qualitatively different and calculate the extrinsic PA versus  the intrinsic DA phonon contributions to the relaxation rate. For a typical doping level in experiments, $n=\bar{n}\times10^{12}$ cm$^{-2}$ with $\bar{n}\sim 1$, we have $\e_{F}\sim 1350$ K and for energies and temperatures up to room temperature, the system of massless Dirac Fermions is a well defined Fermi gas where carrier scattering events due to PA and DA phonons are kinematically quasielastic, {\it i.e.} $\left| \e-\e_{F} \right|,~ T\ll \e_{F}$. 
Therefore, we consider three typical temperature regions with boundaries given in terms of two different BG temperatures, $T^{PA, DA}_{BG}$. Because $T^{DA}_{BG}/T^{PA}_{BG}\approx 8$, all three regions are well defined.

In the high temperature regime, $T\gg T^{PA, DA}_{BG}$, PA and DA phonons with energies $\o_{s \bf q}\sim T^{s}_{BG}$ are important in the relaxation processes and because  $\o_{s \bf q}\ll T$, such scattering events are called \cite{GL} statically quasielastic. Under this severe condition, we have $y\ll 1$ so that the phonon Bose factors are large, $N(y)\approx 1/y\gg 1$, and scattering is dominated by induced phonon emission and absorption processes. Therefore, in Eq.~(\ref{MRRF}) to leading order in $y$, we can replace $f(x\pm y)$ by $f(x)$ while $N(y) + \frac{1}{2} \pm \frac{1}{2} $ by $1/y$. Then, ${\cal G}_{2+m}(x)\approx b T/T_{BG}$ (with $m=0$ and $b=1/3$ for PA and $m=1$ and $b=\pi/16$ for DA phonons) and for the momentum relaxation rate in this regime, we obtain 
\begin{eqnarray}\label{MRREL}
{1\over\tau^{PA,DA}_{1}(\e_{\bf k})}&=&\frac{b c^{2}}{2\pi}\frac{1}{{\bar \tau}_{PA,DA}}\left(\frac{k_{F}}{p_{0}}\right)^{m+1}\frac{T}{T_{F}}~.
\end{eqnarray}
Here the relaxation rate due to DA phonons reproduce the previous results from Refs.~\onlinecite{Hwang2007,Kaasbjerg2012}. The linear $T$ dependence of the momentum relaxation rate both due to PA and DA phonon scattering and its independence on the energy $\e-\e_{F}$ of a test electron are distinctive features of statically quasielastic electron-phonon scattering in the high $T$ regime where the scale of variation of $\tau_{1}(\e)$ is $T_{F}$, which is larger than $T^{s}_{BG}$, and the Pauli exclusion principle does not play an essential role. 
In this regime the extrinsic PA and intrinsic DA contributions to the momentum relaxation rate differ by a factor
$\tau^{DA}_{1}(\e_{\bf k})/\tau^{PA}_{1}(\e_{\bf k})=\gamma_{H} p_{0}/k_{F} \approx \sqrt{0.5/\bar{n}}$
where $\gamma_{H}=\frac{b_{PA}c^{2}_{PA}\bar{\tau}_{DA}}{b_{DA}c^{2}_{DA}\bar{\tau}_{PA}}\approx 0.5$ is determined by the elastic properties of graphene and GaAs. Thus, in this regime we find that independent of $\e-\e_{F}$ and $T$, the extrinsic PA phonon contribution dominates the intrinsic DA phonon contribution to the relaxation rate at densities smaller than $n=5\times 10^{11}$ cm$^{-2}$ and this enhancement is a square-root function with $n$.

\begin{figure}[t]
\includegraphics[width=0.49\linewidth]{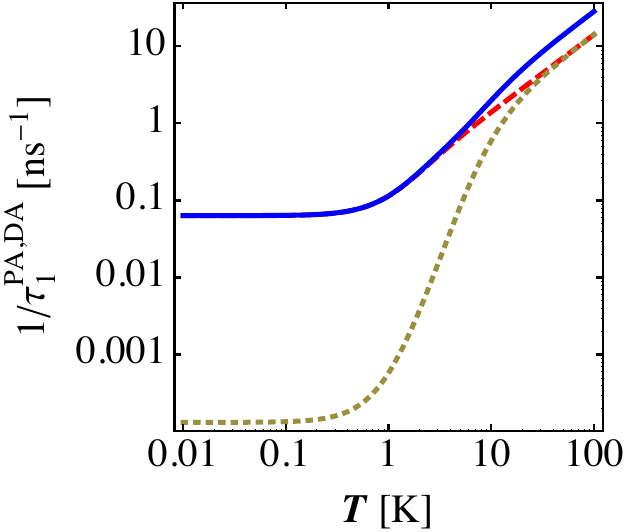}\includegraphics[width=0.49\linewidth,height=0.42\linewidth]{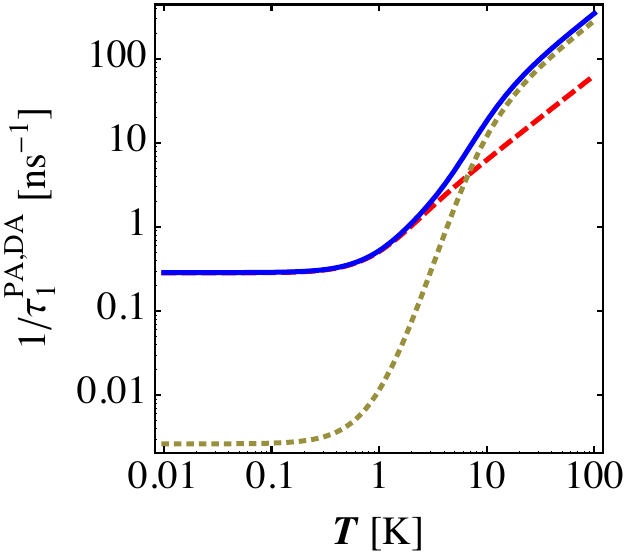}\\
\includegraphics[width=0.49\linewidth]{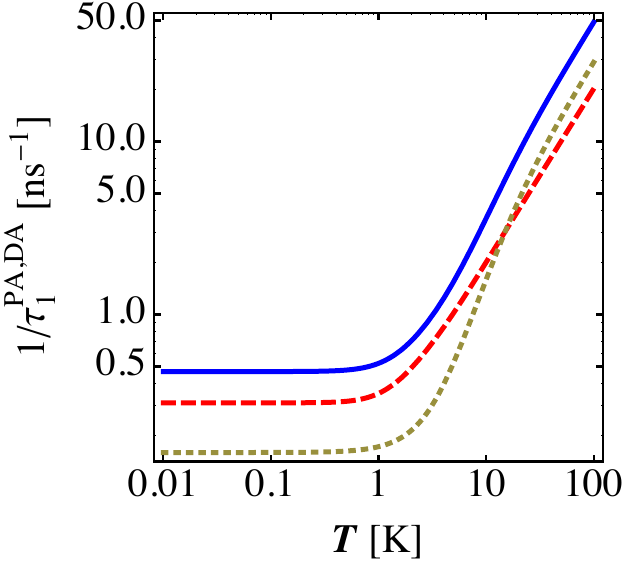}\includegraphics[width=0.49\linewidth]{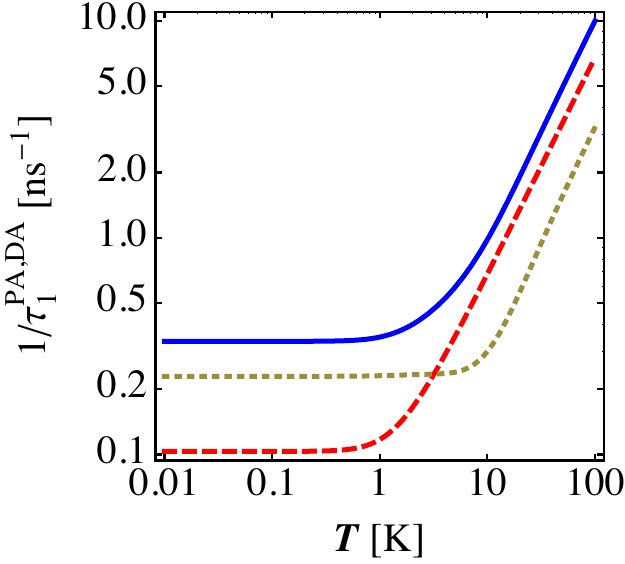}
\caption{The electron momentum relaxation rate $1/\tau_1^{PA,DA}(\e_k)$ as a function of temperature for different values of the electron energy and density. The dashed and dotted lines correspond to the extrinsic PA and intrinsic DA phonon scattering mechanisms. The solid curve is the total rate. Four different situations are presented: (top, left) $\e_k-\e_F=3$ K and $n=0.5\times10^{12}$ cm$^{-2}$, (top, right) $\e_k-\e_F=3$ K and $n=10\times10^{12}$ cm$^{-2}$, (bottom, left) $\e_k-\e_F=15$ K and $n=1.0\times10^{12}$ cm$^{-2}$, and (bottom, right) $\e_k-\e_F=30$ K and $n=0.1\times10^{12}$ cm$^{-2}$.} 
\label{fig1}
\end{figure}

In the low temperature regime, $T\ll T^{PA, DA}_{BG}$, if additionally the energy of a test electron is small, $|\e-\e_{F}|\ll T^{PA, DA}_{BG} $, electronic transitions are dominated by small angle scattering events. The typical phonon momenta, $q\ll k_{F}$, correspond to the $z\rightarrow 0$ limit and one can extend the integration over $z$ in Eq.~(\ref{MRRF}) up to infinity. Replacing also $1-z^{2}$ by $1$ and taking into account that $z\approx y\left(T/T_{BG} \right)$, we have ${\cal G}_{k}(x)\approx \left(T/T_{BG}\right)^{k+1} {\cal F}_{k}(x)$ where we define the function ${\cal F}_{k}(x)=\sum_{\pm}\int^{\infty}_{0} dy y^{k}\Psi^{\pm}(x,y)$. Hence, the momentum relaxation rate in the low $T$ regime is given as
\begin{eqnarray}\label{MRRSAS}
{1\over\tau^{PA,DA}_{1}(\e_{\bf k})}&=& \frac{a c^{2}}{\pi} \frac{1}{{\bar \tau}_{PA,DA}}
\left(\frac{k_{F}}{p_{0}}\right)^{m+1}  \left(\frac{T}{T^{PA,DA}_{BG}}\right)^{m+3} \\
&\times&  {\cal F}_{2+m}(x) \nn
\end{eqnarray}
Here one should distinguish two subregimes. For thermal electrons $|\e-\e_{F}|\lesssim T$, electronic transitions with typical phonon momenta $\hbar q\sim T/v_{PA,DA}$ dominate. The function ${\cal F}_{k}(x)$ weekly depends on $x$ for $x\lesssim 1$ and can be replaced by ${\cal F}_{k}(0)$. Hence in this BG regime the momentum relaxation rate (\ref{MRRSAS}) exhibits $T^{3}$ and $T^{4}$ dependences, respectively, for PA and DA phonon scattering. 
For hot electrons in the opposite $T\rightarrow 0$ limit, electronic transitions are dominated with spontaneous emission of phonons. We have ${\cal F}_{k}(x)\approx |x|^{k}/(1+k)$ for $|x|\gg1$ and the relaxation rate (\ref{MRRSAS}) does not depend on $T$ and is proportional to $\left(|\e-\e_{F}|/T^{PA, DA}_{BG}\right)^{m+3}$.
The direct comparison of the extrinsic PA and intrinsic DA contributions to the momentum relaxation rate in the low $T$ regime gives $\tau^{DA}_{1}(\e_{\bf k})/\tau^{PA}_{1}(\e_{\bf k})=\gamma_{L}\left(p_{0}/k_{F}\right) \left(T_{PA}/T\right)$ for thermal electrons
where the factor $\gamma_{L}=\frac{a^{3}_{DA}c^{2}_{PA}{\cal F}_{2}(0)\bar{\tau}_{DA}}{a^{3}_{PA}c^{2}_{DA}{\cal F}_{3}(0)\bar{\tau}_{PA}}\approx 41$. For hot electrons $\gamma_{L}$ is even larger. Thus, in the low $T$ regime, the intrinsic DA contribution is negligible while the extrinsic PA contribution leads not only to strong enhancement of the momentum relaxation rate but changes its dependence on the lattice temperature from $T^{4}$ to $T^{3}$.

\begin{figure}[t]
\includegraphics[width=0.49\linewidth]{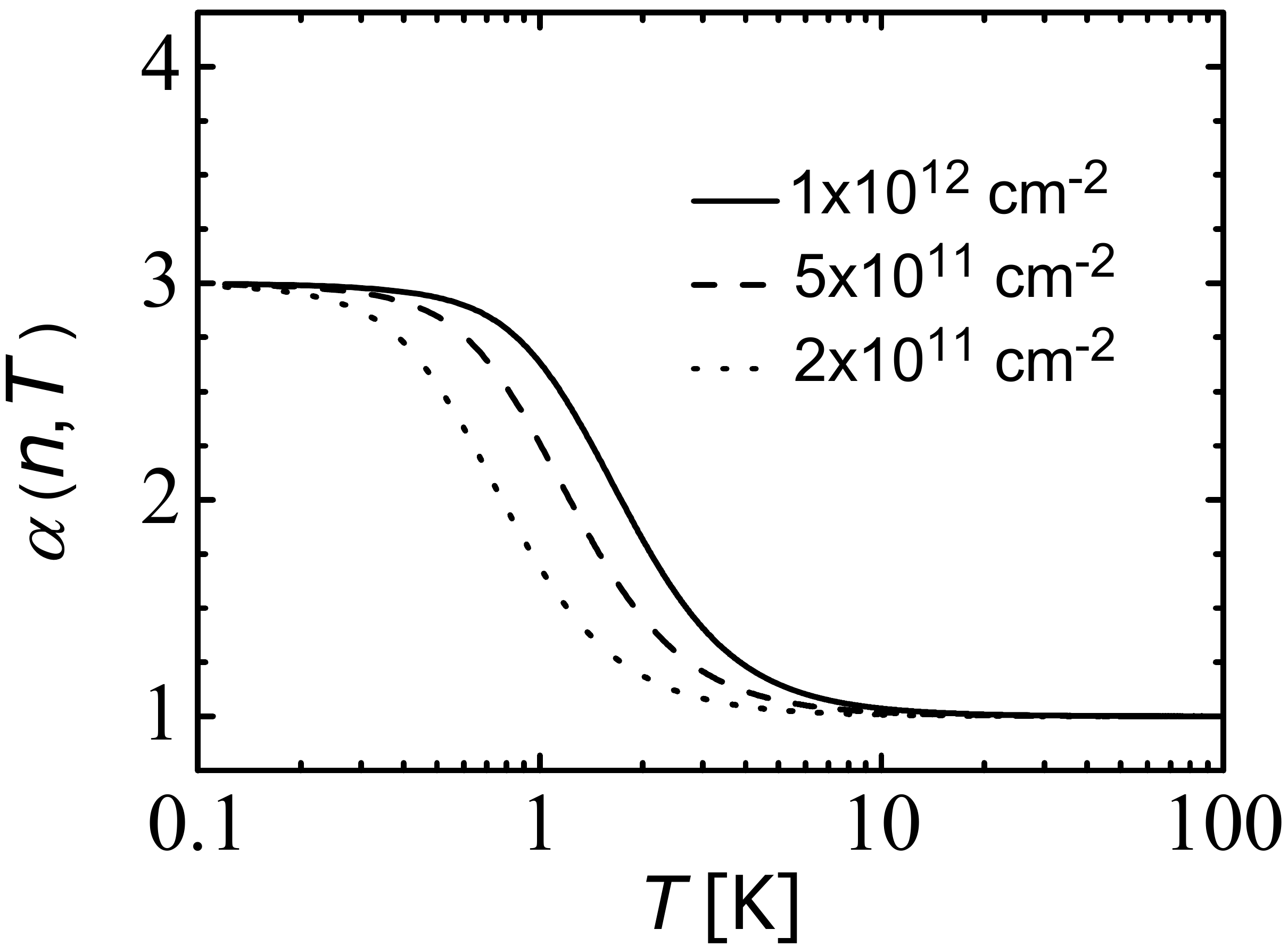}\includegraphics[width=0.49\linewidth]{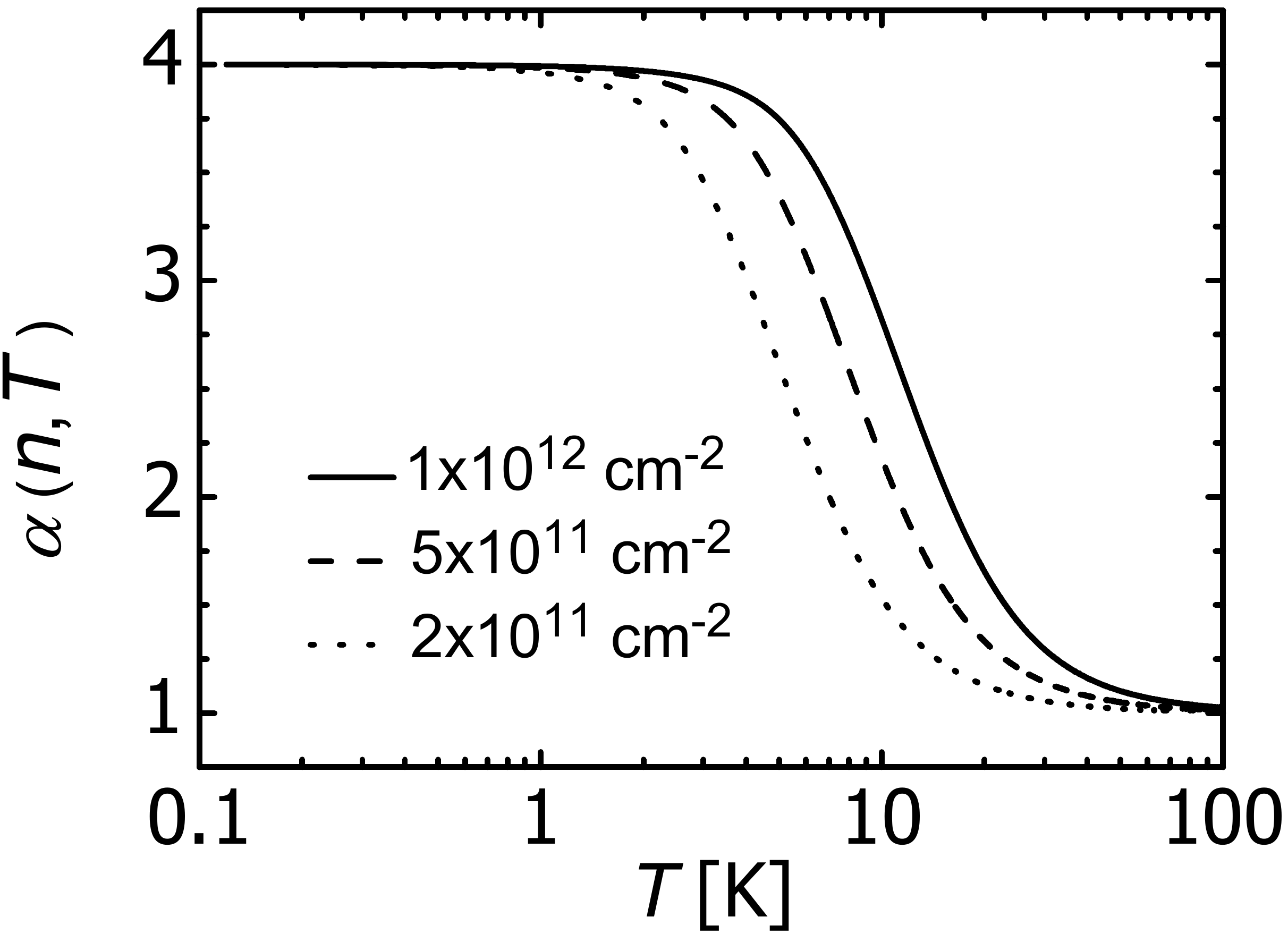}\\
\includegraphics[width=0.49\linewidth]{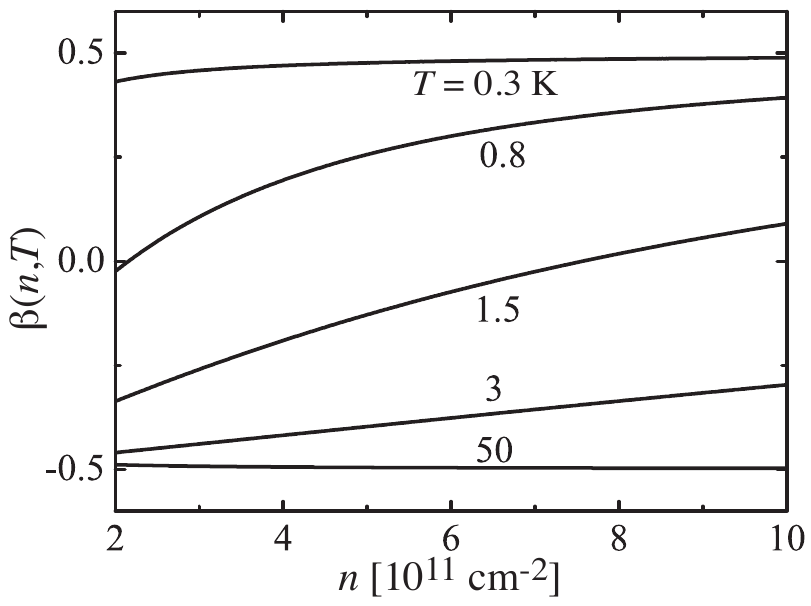}\includegraphics[width=0.49\linewidth]{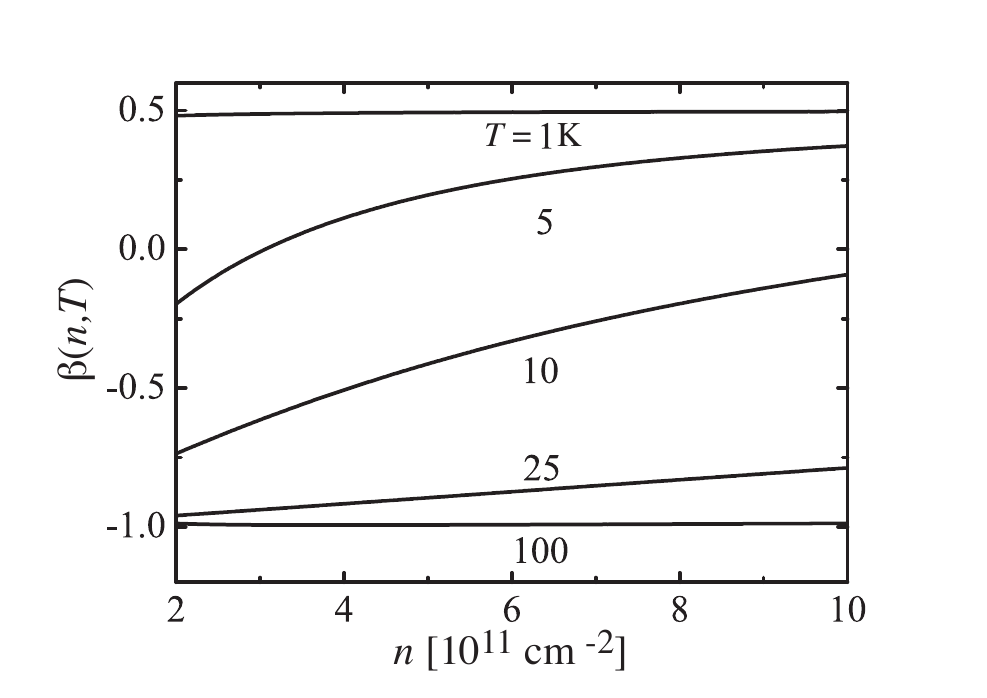}
\caption{The exponents $\alpha(n,T)$ (top panels) and $\beta(n,T)$ (bottom panels) describing the mobility behavior $\mu_{s}\propto T^{-\alpha(n,T)}$ and $\mu_{s}\propto n^{\beta(n,T)}$, respectively, as a function of $T$ and $n$. The panels on the left and right hand sides depict, respectively, the PA and DA phonon limited mobilities $\mu_{PA}$ and $\mu_{DA}$. }
\label{fig2}
\end{figure}

In the region of intermediate temperatures, $T^{PA}_{BG} \ll T \ll T^{DA}_{BG}$, small angle scattering events dominate the DA phonon relaxation with the rate given by  Eq.~(\ref{MRRSAS}) while the PA phonon relaxation is governed by quasielastic large angle scattering events with the rate (\ref{MRREL}). Therefore, the momentum relaxation of thermal electrons due to scattering from intrinsic DA phonons is still suppressed in comparison with that from extrinsic PA phonons by a factor $\tau^{DA}_{1}(\e_{\bf k})/\tau^{PA}_{1}(\e_{\bf k})=\gamma_{I}\left(p_{0}/k_{F}\right)  \left(T^{DA}_{BG}/T\right)^{3}$
where $\gamma_{I}=\frac{b_{PA} c^{2}_{PA}\bar{\tau}_{DA}}{{\cal F}_{3}(0) c^{2}_{DA} \bar{\tau}_{PA}}$. As seen this suppression depends strongly on $T$, however, due to the small numerical pre-factor, $\gamma_{I} \approx 4\times 10^{-3}$ ($2.5 \times 10^{-2}$ for hot electrons), it is large only near the lower edge, $T\gtrsim T_{BG}^{PA}$, of this intermediate temperature region with the relaxation rate linear in $T$. Towards the upper edge, $T\lesssim T_{BG}^{DA}$, the DA phonon contribution to the momentum relaxation rate increases and at high densities the PA contribution prevails with the $T^{4}$ behavior of the relaxation rate.


Within the Boltzmann transport theory, the electron momentum relaxation time averaged over its energy $\e$, $\tau^{s}_{1}(n,T)=\int d\e D(\e) \tau^{s}_{1}(\e) \left(-\partial f(\e)/\partial \e \right)$ ($D(\e)$ is the electron density of states), determines the carrier mobility in graphene samples as \cite{Nomura20067}
\begin{equation}\label{MOB}
\mu_{PA,DA}(n,T)= \frac{e v_{F}}{\hbar k_{F}} \tau^{PA,DA}_{1}(n,T)~.
\end{equation}  
As seen from Eqs.~(\ref{MRREL}) and ~(\ref{MRRSAS}) in the high $T$ regime as well as for thermal electrons in the low $T$ BG regime, the momentum relaxation rate of a test electron is independent of its energy. Therefore, we find that in the high $T$ regime the mobility shows the same, $\mu \propto T^{-1}$, temperature dependence for both the PA and DA scattering mechanisms but different density dependences,  $\mu \propto 1/\sqrt{n}$ and $\mu \propto 1/n$, respectively, for PA and DA phonon scattering. In the low $T$ BG regime the mobility exhibits the same density dependence, $\mu \propto \sqrt{n}$, but different temperature dependences, $\mu \propto T^{-3}$ and $\mu \propto T^{-4}$, respectively, for PA and DA phonon scattering.

\begin{figure}[t]
\includegraphics[width=0.49\linewidth]{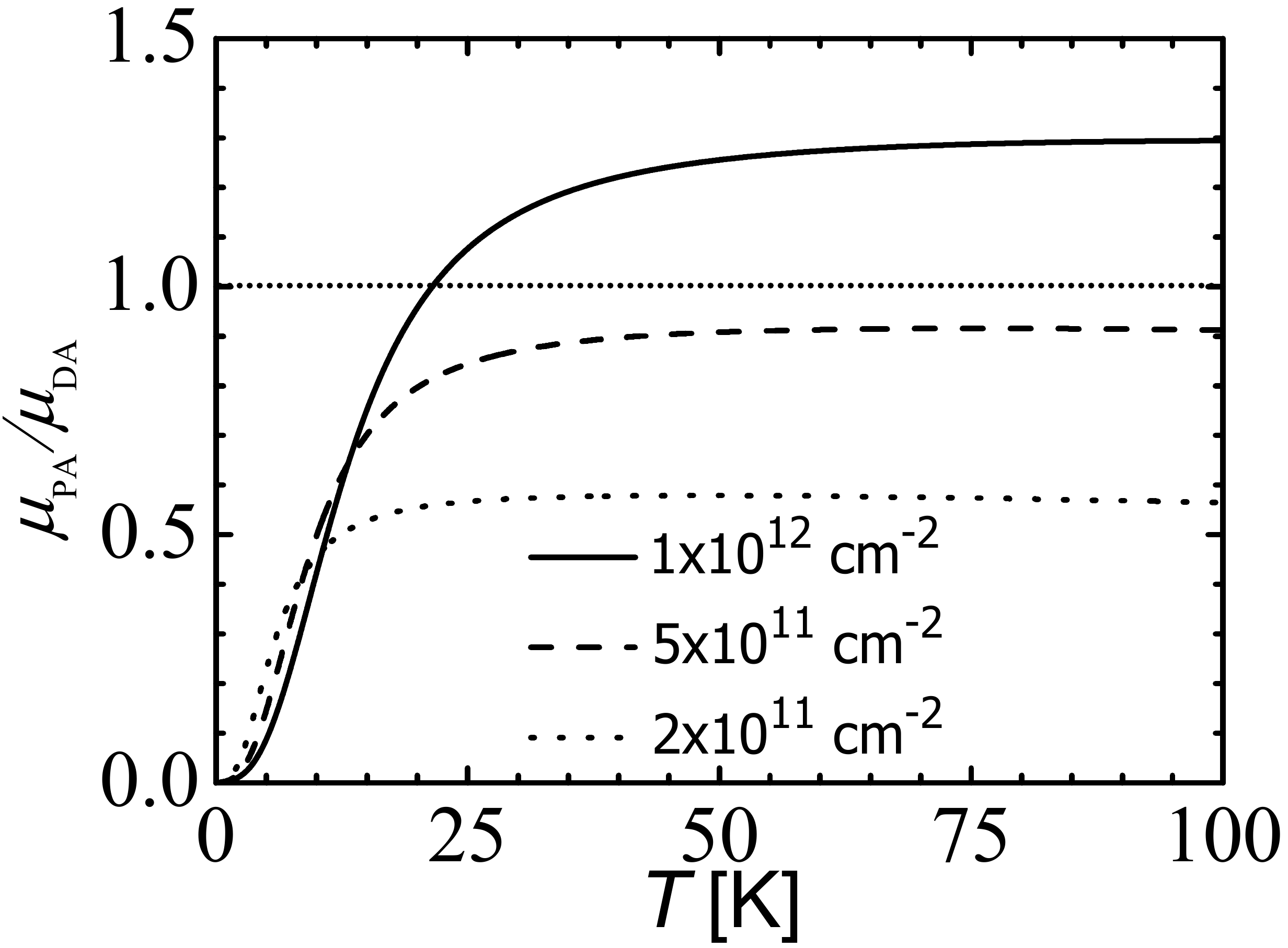}\includegraphics[width=0.49\linewidth]{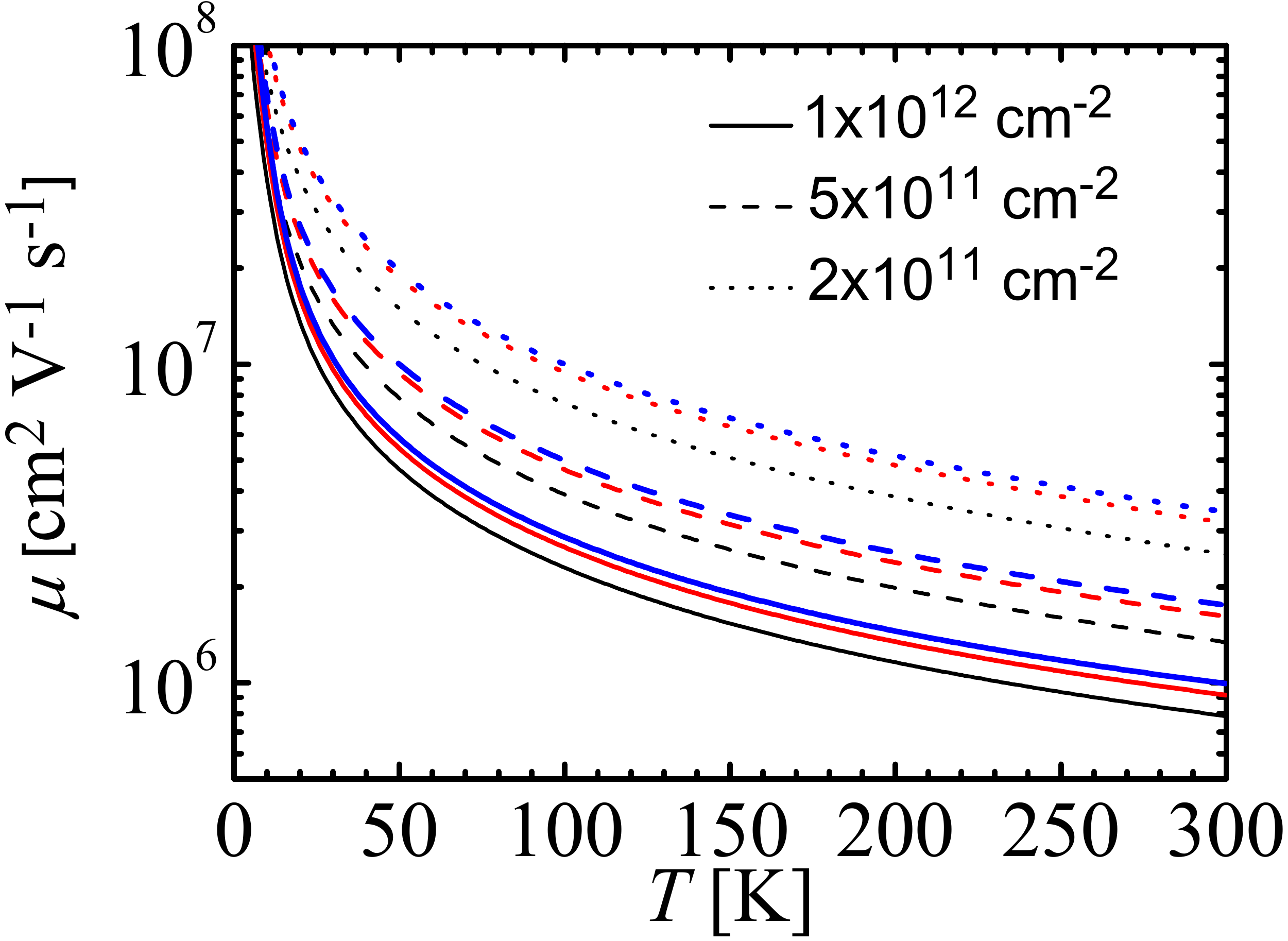}
\caption{(left) The mobility ratio $\mu_{PA}/\mu_{DA}$ and (right) the total mobility $\mu=\mu_{PA}\mu_{DA}/(\mu_{PA}+\mu_{DA})$ versus  temperature. The solid, dashed, and dotted curves correspond to the electron density $n=10$, $5$ and $2\times 10^{11}$ cm$^{-2}$. In the right panel, the curves in each set are calculated in three different approximations for the PA phonon potential. The lower curves refer to the $\left(\hat{q}_{x}\hat{q}_{x}\right)^{2}\approx 1/4$ and $d=0$ approximation, the middle curves take into account the angle-dependence of the PA vertex with $d=0$, the upper curves include both the effect of anisotropy and of the finite distance with $d=5$ \AA.}
\label{fig3}
\end{figure}

In Figs.~\ref{fig1}-\ref{fig3} we present our numerical calculations based on Eqs.~(\ref{MRREX}), (\ref{MRRF}) and (\ref{MOB}). 
In Fig.~\ref{fig1} we study the electron momentum relaxation scattered due to the combined action of extrinsic PA and intrinsic DA phonons. The relaxation rate versus  $T$ is plotted for different values of $\e-\e_{F}$ and $n$. It is seen in all panels that at low temperatures the relaxation rate is almost independent of $T$, which corresponds to the behavior of hot electrons with $\e-\e_{F}\gg T$. In the small angle scattering subregime, $T\ll \e-\e_{F}\ll T^{PA,DA}_{BG}$, the intrinsic DA scattering is strongly suppressed with respect to the extrinsic PA scattering ({\it cf.} the (top left) and (top right) panels) and this is in agreement with the above discussion that $1/\tau_{1}^{PA,DA}\propto \left(\e-\e_{F}/T^{PA,DA}_{BG}\right)^{3+m}$ in this subregime. With an increase of $\e-\e_{F}$ the dependence weakens and in the large angle scattering subregime, $\e-\e_{F}\gtrsim T^{PA,DA}_{BG}$, the PA and DA phonon contributions to the momentum relaxation of hot electrons become of the same order and the DA phonon contribution can even dominate at low densities ({\it cf.} the (bottom right) panel). 
In the high $T$ regime, as seen in all panels of Fig.~\ref{fig1} the momentum relaxation rate exhibits the same linear $T$ dependence, in agreement with Eq.~(\ref{MRREL}). 
The PA and DA phonon mechanisms make equal contributions to the relaxation rate at carrier densities $n=5.0\times10^{11}$ cm$^{-2}$ ({\it cf.} (top left) panel). In this regime the relaxation rate is independent of $\e-\e_{F}$ and the relative PA and DA contribution can be tuned by changing solely the density $n$. At intermediate temperatures the momentum relaxation shows a crossover from the $T$ independent to the linear $T$ regime. The total rate depending on $n$ and $T$ can be dominated either by PA phonon scattering with the $T^{3}$ behavior for smaller values of $n$ or by DA phonon scattering with the stronger $T^{4}$ dependence for higher values of $n$ ({\it cf.} (top right) panel).

In Fig.~\ref{fig2} we plot the exponents $\alpha(n,T)$ and $\beta(n,T)$, {\it i.e} $\mu_{PA,DA}\propto T^{-\alpha(n,T)}$ and $\mu_{PA,DA}\propto n^{\beta(n,T)}$, respectively, as a function of $T$ and $n$. 
It is seen that the temperature and density behavior of the mobility calculated numerically for the PA and DA scattering mechanisms is consistent with the above analytical findings. For any value of $n$, the $T^{-1}$ and $T^{-4}$ dependence of the mobility, respectively for PA and DA phonon scattering, occurs in a wider temperature range than the $T^{-1}$ and $T^{-3}$ dependence for DA and PA scattering. Such a reversed behavior stems from the substantial difference of the sound velocity in GaAs and graphene and can be used to distinguish between the extrinsic PA and extrinsic DA mechanisms of the momentum relaxation.
As seen from the bottom panels in Fig.~\ref{fig2} at low $T$ the exponent $\beta(n,T) \approx 1/2 $ for both the PA and DA mechanisms. With an increase of $T$ the PA contribution to the mobility shows a crossover to the behavior with $\beta(n,T) \approx -1/2 $ while the DA contribution exhibits a crossover to the behavior with $\beta(n,T) \approx -1$.

In Fig.~\ref{fig3} we plot the temperature dependence of the relative and combined contributions to the mobility made by the PA to DA phonon scattering for different values of the electron density. It is seen from the left panel that at densities $n< 5\times10^{11}$ cm$^{-2}$, the ratio $\mu_{PA}/\mu_{DA}< 1$ so that PA phonon scattering is the dominant mechanism, limiting the mobility both at low and high temperatures.  
The right panel shows the total mobility versus temperature, which we obtain applying the Matthiessen rule, ${1/\mu}={1/\mu_{PA}}+{1/\mu_{DA}}$. Here we calculate the PA contribution to the mobility including both the angle-dependence of the piezoelectric potential and the finite distance $d$ between the graphene sheet and the substrate. From the comparison of the upper thick curves with the lower two curves in each set of lines, it is seen explicitly that the effect of the finite distance and the anisotropy of the PA potential is weak. 

In conclusion, the piezoelectric potential of acoustical phonons propagating on the surface of the GaAs substrate is an important factor in limiting the mobility of Dirac fermions. At low densities PA phonon scattering is the dominant momentum relaxation mechanism in graphene. At high temperatures it changes qualitatively the density dependence of the mobility while in the Bloch-Gr\"uneisen regime the power law dependence on temperature.

This work was supported by the ESF-Eurocores program EuroGRAPHENE (CONGRAN project) and the Vlemish Science Foundation (FWO-Vl). One of us (S.M.B.) acknowledges the Belgian Science Policy (BELSPO and EU).

\end{document}